\documentclass{aa}

\usepackage{graphicx}
\usepackage{txfonts}
\usepackage{booktabs}
\usepackage{multirow}

\begin{document}

   \title{A 2022 $\tau$-Herculid meteor cluster from an airborne experiment: automated detection, characterization, and consequences for meteoroids}

   
   \titlerunning{$\tau$-Herculid meteor cluster}

   \author{J. Vaubaillon \inst{1} \and
           C. Loir       \inst{1} \and
          C. Ciocan      \inst{2} \and
          M. Kandeepan   \inst{2} \and
          M. Millet      \inst{2} \and
          A. Cassagne    \inst{2} \and
          L. Lacassagne  \inst{2} \and\\
          P. Da Fonseca  \inst{1} \and
          F. Zander      \inst{3} \and
          D. Buttsworth  \inst{3} \and
          S. Loehle      \inst{4} \and
          J. Toth        \inst{5} \and
          S. Gray        \inst{6} \and
          A. Moingeon    \inst{1} \and
          N. Rambaux     \inst{1}
          }

   \institute{IMCCE, CNRS, Observatoire de Paris, PSL Universit\'e, Sorbonne Universit\'e, Universit\'e de Lille 1, UMR 8028 du CNRS, 77 av. Denfert-Rochereau 75014 Paris, France  \quad \email{vaubaill@imcce.fr}
         \and
         Sorbonne Université, CNRS, LIP6, F-75005 Paris, France \quad \email{lionel.lacassagne@lip6.fr}
         \and
             University of Southern Queensland, West St, Toowoomba, 4350, Australia
        \and
            High Enthalpy Flow Diagnostics Group (HEFDiG), Institute of Space Systems, University of Stuttgart, Pfaffenwaldring 29, 70569 Stuttgart, Germany
        \and
            Faculty of Mathematics, Physics and Informatics, Comenius University in Bratislava, Slovakia
        \and
            Rocket Technologies International, Seventeen Mile Rd, Helidon, 4344, Australia
             }

   \date{Received xx 2022; accepted XXX 2022}

 
  \abstract
   {The existence of meteor clusters has long since been a subject of speculation and so far only seven events have been reported, among which two involve less than five meteors, and three were seen during the Leonid storms.}
   {The 1995 outburst of Comet 73P/Schwassmann-Wachmann was predicted to result in a meteor shower in May 2022. We detected the shower, proved this to be the result of this outburst, and detected another meteor cluster during the same observation mission.}
   {The  $\tau$-Herculids meteor shower outburst on 31 May 2022 was continuously monitored for 4 hours during an airborne campaign. The video data were analyzed using a recently developed computer-vision processing chain for meteor real-time detection.}
   {We report and characterize the detection of a meteor cluster involving 38 fragments, detected at 06:48 UT for a total duration of 11.3 s.
   The derived cumulative size frequency distribution index is relatively shallow: $s=3.1$.
   Our open-source computer-vision processing chain (named FMDT) detects 100\% of the meteors that a human eye is able to detect in the video. Classical automated motion detection assuming a static camera was not suitable for the stabilized camera setup because of residual motion.
   }
   {From all reported meteor clusters, we crudely estimate their occurrence to be less than one per million observed meteors.
   Low heliocentric distance enhances the probability of such meteoroid self-disruption in the interplanetary space.}

   \keywords{meteoroids --
            Comets: general --
            method: data analysis
               }

   \maketitle
%

\section{Introduction}

Meteor clusters are the occurrence of many (typically more than three) meteors detected in a restricted portion of the sky within a few seconds \citep[typically less than 5s; for a review see][and references]{Koten2017}.
The existence of meteor clusters has been suspected for decades, but only a handful (namely six) of observations have been reported \citep{Koten2017}.
In particular, the Leonid storms occurring between 1998 and 2002 raised the question as to whether or not such observations happened simply by chance, given the high number of meteors recorded during each event.
Evidence of their genuine existence has been reported that disfavors  their observation simply being the result of statistical fluctuations \citep{Watanabe2002,Toth2004a}.
Given the ever growing number of meteor cameras surveying the sky around the globe every night, such events are expected to be more frequently reported.
However, even with more than a thousand meteor detection cameras running today \citep[see][for a review of all the networks]{Koten2019}, meteor cluster observations are still very rare events.

The exact origin of meteor clusters is poorly known.
A probable process is thermal stressing of very fragile comet dust \citep{Watanabe2003}, a hypothesis recently confirmed by \cite{Capek2022} for the case of the 2016 September $\epsilon$- Perseid (SPE) cluster. The meteoroid disruption drives the level of meteor showers, which itself depends on the structure of the comet.
\cite{Jenniskens2008} reported a lack of fluffy meteoroid in an old Leonid trail, which these authors suggested is possibly explained by meteoroid self-disruption in the inter-planetary space (although, other hypotheses might explain this observation).
In order to explain the present quasi-steady-state of the level of sporadic meteors and of the amount of zodiacal dust, models must take the meteoroid life expectancy into account as well as the replenishment mechanism \citep{Wiegert2009,Levasseur2020}.
Such mechanisms include the gravitational perturbation of long-period comets, the structure and population of the Oort cloud, the role of giant planets (especially Jupiter) in removing or accreting small bodies in the inner Solar System, and so on.
Therefore, the frequency of meteoroid self-fragmentation in space has implications for our current understanding of the Solar System.

Since \cite{Koten2017}, only one meteor cluster observation has been reported \citep{HawaiiCluster2021}, although an extensive search was recently performed among the Geminids \citep{Koten2021}.
One open question refers to the frequency of spontaneous meteoroid breakup in interplanetary space, which would lead to meteor clusters and how these breakups would influence the lifetime expectancy of meteoroids.
Here, we report another unambiguous detection of a meteor cluster of 34 fragments, detected within 7.5 s during the 2022 $\tau$-Herculids outburst caused by the 1995 trail ejected from Jupiter family comet 73P/Schwassmann-Wachmann 3 (hereafter 73P). The detection was realized using a novel computer-vision application that was able to detect 100\,\% of the meteors that a human eye can see in the video. The results enable a discussion of the origin, frequency, and implications of such events.

\section{Observations}\label{sec:obs}

\begin{figure*}[!htbp]
\centering
\includegraphics[width=0.9\textwidth,keepaspectratio]{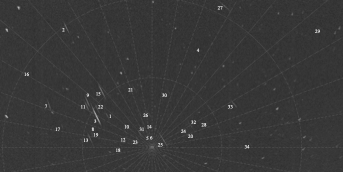}
\caption{Composite closeup view of the detection of the 34 $\tau$-Herculids meteor cluster from an airborne observation campaign. The center of the field of view points toward the little dipper (\textit{Ursa Minor}) constellation.}
\label{fig:composite}
\end{figure*}

\newcommand{\MR}[1]{\multirow{2}{*}{#1}}

\begin{table*}
\centering
\caption{$\tau$-Herculid meteor cluster characterization: beginning time in UT, duration ($\mathcal{D}$) in seconds, and apparent magnitude ($\mathcal{M}$). Both ground-truth and automatic detection results (see Sec.~\ref{sec:detection}) are reported.}
{\small
\begin{tabular}{rrrrrr|rrrrrr}
\toprule
& \multicolumn{2}{c}{Ground truth} & \multicolumn{2}{c}{Automatic detection} & & & \multicolumn{2}{c}{Ground truth} & \multicolumn{2}{c}{Automatic detection} & \\
\cmidrule(lr){2-3} \cmidrule(lr){4-5} \cmidrule(lr){8-9} \cmidrule(lr){10-11}
{\#} & {Beginning} & {$\mathcal{D}$} & {Beginning} & {$\mathcal{D}$} & {$\mathcal{M}$} & {\#} & {Beginning} & {$\mathcal{D}$} & {Beginning} & {$\mathcal{D}$} & {$\mathcal{M}$} \\
\midrule
  1 & {06:48:55.959} & 0.30 & {06:48:55.959} & 0.30 & -0.19 &     19  &     {06:48:59.009}  &     0.20  & {06:48:59.059} & 0.15 &     0.62  \\ 
  2 & {06:48:56.359} & 0.50 & {06:48:56.409} & 0.35 & -1.50 &     20  &     {06:48:59.209}  &     0.10  & {06:48:59.209} & 0.10 &     1.79  \\ 
  3 & {06:48:56.359} & 0.80 & {06:48:56.359} & 0.80 & -2.06 &     21  &     {06:48:59.359}  &     0.25  & {06:48:59.409} & 0.20 &    -0.21  \\ 
  4 & {06:48:56.909} & 0.10 & {06:48:56.909} & 0.10 &  0.87 &     22  &     {06:48:59.559}  &     0.35  & {06:48:59.559} & 0.30 &    -0.56  \\ 
  5 & {06:48:57.209} & 0.10 & {06:48:57.209} & 0.10 &  1.19 &     23  &     {06:48:59.759}  &     0.35  & {06:48:59.759} & 0.10 &     0.73  \\ 
  6 & {06:48:57.259} & 0.20 & {06:48:57.309} & 0.10 &  0.92 &     24  &     {06:48:59.809}  &     0.15  & {06:48:59.809} & 0.10 &     1.33  \\ 
  7 & {06:48:57.509} & 0.45 & {06:48:57.509} & 0.40 & -1.07 &     25  &     {06:48:59.809}  &     0.50  & {06:48:59.809} & 0.50 &    -0.65  \\ 
  8 & {06:48:57.559} & 0.15 & {06:48:57.599} & 0.15 &  0.86 &     26  &     {06:48:59.859}  &     0.25  & {06:48:59.859} & 0.20 &     0.41  \\ 
  9 & {06:48:57.559} & 0.45 & {06:48:57.599} & 0.25 & -1.40 &     27  &     {06:49:00.009}  &     0.35  & {06:49:00.009} & 0.30 &    -1.72  \\ 
 10 & {06:48:57.609} & 0.15 & {06:48:57.659} & 0.10 &  0.27 &     28  &     {06:49:00.559}  &     0.15  & {06:49:00.559} & 0.15 &     0.70  \\ 
 11 & {06:48:57.709} & 0.45 & {06:48:57.809} & 0.25 & -0.88 &     29  &     {06:49:00.609}  &     0.30  & {06:49:00.709} & 0.10 &    -1.09  \\ 
 12 & {06:48:57.809} & 0.20 & {06:48:57.809} & 0.15 &  0.92 & \MR{30} & \MR{{06:49:00.709}} & \MR{0.25} & {06:49:00.759} & 0.10 & \MR{0.26} \\ 
 13 & {06:48:57.859} & 0.50 & {06:48:57.859} & 0.50 & -0.68 &         &                     &           & {06:49:00.809} & 0.15 &           \\ 
 14 & {06:48:58.159} & 0.15 & {06:48:58.159} & 0.15 &  1.44 &     31  &     {06:49:00.809}  &     0.35  & {06:49:00.859} & 0.25 &    -0.31  \\ 
 15 & {06:48:58.559} & 0.55 & {06:48:58.659} & 0.45 & -1.32 &     32  &     {06:49:02.009}  &     0.30  & {06:49:02.009} & 0.30 &     0.27  \\ 
 16 & {06:48:58.609} & 0.20 & {06:48:58.659} & 0.10 & -0.31 &     33  &     {06:49:02.059}  &     0.25  & {06:49:02.059} & 0.20 &    -0.82  \\ 
 17 & {06:48:58.709} & 0.30 & {06:48:58.709} & 0.25 & -0.01 &     34  &     {06:49:03.309}  &     0.15  & {06:49:03.309} & 0.15 &     0.56  \\ 
 18 & {06:48:58.859} & 0.15 & {06:48:58.859} & 0.15 &  0.85 &         &                     &           &                &      &           \\ 
\bottomrule
\end{tabular}
}
\label{tab:charac}
\end{table*}

\subsection{Campaign and instrument}\label{sec:instr}

\citet{YeVaubaillon2022} predicted that the 2022 $\tau$-Herculids meteor shower would be visible on 31 May 2022, and indeed it was successfully observed during an airborne observation campaign led by the University of Southern Queensland (F.Z., D.B.) and supported by Rocket Technologies International (S.G.).
On board the aircraft, a suite of low-light scientific cameras were installed at several windows, and the sky was monitored continuously.
A detailed description of the whole campaign is beyond the scope of this paper, but will be published in a dedicated paper.
In addition to the imaging systems, spectroscopic systems were mounted in parallel. Windows on both sides of the aircraft were equipped with cameras, and the flight path was chosen according to the predictions \citep{YeVaubaillon2022}.

In this paper, we present results from data collected by a Basler acA1920-155um camera equipped with a Basler 6mm f/1.4 lens.
The gain was set at maximum value (36) and 20 images per second were taken during the 4 hours of the flight.
In order to compensate for the airplane (Phenom 300) roll motion, a G6-Max camera stabilizer was used.
The camera was controlled with a RaspBerry-4 mini-computer, running the "RMS" acquisition and meteor detection software \citep{Vida.et.al2016,Vida2021}.
In addition, an AMOS-Spec-HR camera (Comenius Univ.) was mounted at another plane window. The hardware was a DMK 33UX252 (resolution of $2048 \times 1536$ px and set to 14fps) equipped with a 6 mm, F/1.4 lens, providing a FOV of $60 \times 45 \deg$.

\subsection{Detection of the meteor cluster}\label{sec:descri}

With the described settings, we detected 165 $\tau$-Herculids meteors and five sporadic meteors. 
At the time of the shower outburst maximum (around 05:00 UT), we detected about one meteor per minute.
When the level of the shower was decreasing, starting at 06:48:56 UT, the Basler camera detected 34 meteors, all coming from the $\tau$-Herculids radiant. The AMOS camera, being slightly more sensitive, allowed the detection of 38 meteors within 11.3 s.
Figure~\ref{fig:composite} shows a composite image of the meteor cluster (as detected by the Basler camera).
This cluster observation was not reported by any of the ground-based meteor networks.

\section{Meteor cluster characterization}\label{sec:cara}

The whole meteor cluster characterization was performed with the data from the Basler camera and is detailed in Table ~\ref{tab:charac}.
Further characterizations of the whole shower are ongoing.
The total time duration of the event is 11.3 s, but the Basler camera detected it for 7.5 s only.
The maximum angular distance between all the meteors is $\sim 50 \; \deg$.
The average airplane position during the  cluster was lat=34.20 $\deg$, lon=-101.88 $\deg$, alt=14201  m and the camera was pointing towards its left hand side.
The entry velocity was computed using the algorithm developed by \cite{Neslusan1998}.
Given the low entry velocity of the $\tau$-Herculids (12.2 km.s$^{-1}$), it is reasonable to assume an average meteor altitude of $90$ km.
Individual meteor azimuth and elevation (above the horizon) were measured.
The relative apparent angular distance between each fragment is in the range $[0.25;50.2]\;\deg$, the measured elevation is within $[41.2;70.6]\;\deg$, and the physical distance between two fragments is within $[0.4;90.5]$ km.
Adding the total duration of the event, the maximum possible physical distance between all the fragments is $D_m=[227;244]$ km.

Following the methodology of \cite{Koten2017}, we find that, assuming a Poissonian distribution of meteors (this assumption is discussed in Sect. \ref{sec:discuss:freq}), the probability of such clustering by chance is $\sim 5.5 \times 10^{-22}$ at best.
As a result, we consider that the chance observation of this number of meteors during such a short time period is highly improbable and conclude that the disintegration of a parent $\tau$-Herculid meteoroid took place in the interplanetary space.

Assuming a zero ejection velocity for all the fragments, the maximum time between the parent meteoroid disintegration in interplanetary space and the Earth atmosphere entry strongly depends on the considered meteoroid size.
We converted the apparent magnitude into an absolute magnitude (assuming a meteor altitude of $90$ km), then converting this latter into an equivalent photometric mass \citep[using][]{Hughes1995} and radius.
The latter ranges from $7.5$ mm to $22.4$ mm (assuming a density of 2500 kg.m$^{-3}$).
The total mass of the initial meteoroid is estimated to be 1.16 kg.
The maximum age of the cluster is computed using the smallest size, as this is the most sensitive to the solar radiation pressure, and is found to lie in the range $[13.3;13.8]$ days.

\begin{figure}[!htbp]
\centering
\includegraphics[width=0.49\textwidth,keepaspectratio]{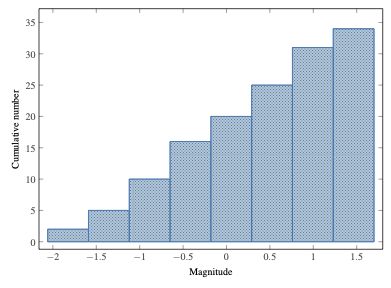}
\caption{Cumulative absolute magnitude distribution of the cluster fragments.}
\label{fig:sfd}
\end{figure}

Figure~\ref{fig:sfd} shows the absolute magnitude distribution of the cluster.
The population index is $r=2.01$, corresponding to a differential size distribution of $s=3.09$.
This feature is discussed in Sect. \ref{sec:discuss:prop}.

\section{Computer-vision detection}\label{sec:detection}

\begin{figure*}[!htbp]
\centering
\includegraphics[width=1\textwidth,keepaspectratio]{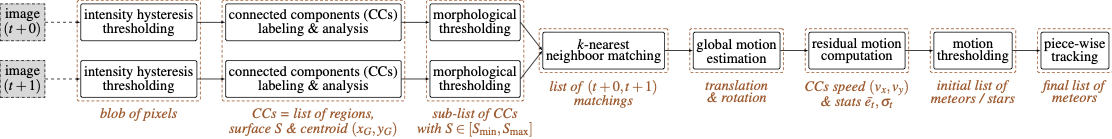}
\caption{Computer-vision detection detailed chain. Plain gray boxes correspond to input data, plain white boxes are the processing, and the italicized brown texts are the processing outputs.}
\label{fig:motion_detection}
\end{figure*}

\begin{figure*}[!htbp]
\centering
\includegraphics[width=0.9\textwidth,keepaspectratio]{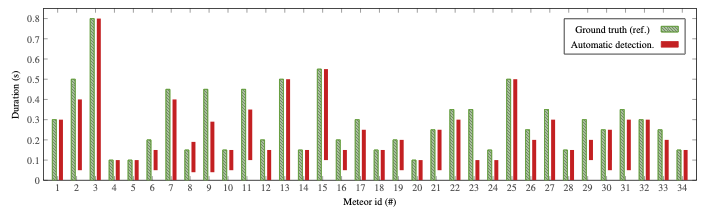}
\label{plot:meteors_duration}
\caption{Overlapping duration of the ground truth and the automatic detection for the meteor cluster. The automatic detection bars are placed relative to the beginning time of each meteor.} 
\end{figure*}

The real-time open-source software detection chain named \emph{Fast Meteor Detection Toolbox} (FMDT) was applied in order to detect the meteor in the imaging data\footnote{FMDT repository: \url{https://github.com/alsoc/fmdt}}. FMDT is derived from software designed to detect meteors on board the ISS or a Cubesat~\citep{Millet2022_Meteorix_COSPAR,Millet2022_Meteorix_WGN,Petreto2018_SIMD_GPU_OF_DASIP}. FMDT is foreseen to be applied to airborne camera systems; for example in atmospheric balloons or aircraft. It is robust to camera movements thanks to a motion-compensation algorithm.

Figure~\ref{fig:motion_detection} presents the whole FMDT detection chain. For each pair of images, an intensity hysteresis threshold, a connected component labeling, and an analysis algorithm~\citep{Lemaitre2020_SIMD_CCL_WPMVP,Lacassagne2009_LSL_ICIP} are applied to get a list of connected components (CCs) with their bounding boxes and surface $S$. Moreover, it also provides the first raw moments to compute the centroid $(x_G,y_G)=(S_x/S,S_y/S)$ of each blob of pixels.  A morphological threshold is then set on the surface $S$ to reject small and big CCs. A $k$-nearest neighbor matching process is then applied to extract pairs of CCs from images $I_{t+0}$ and $I_{t+1}$ with $t$ the image number in the video sequence. These matches are used to perform a first global motion estimation (rigid registration).

This motion estimation is used to classify the CCs into two classes, namely nonmoving stars and moving meteors, according to the following criterion: $|e_k-\bar{e_t}| > \sigma_t$ where $e_k$ is the compensation error of the CC number $k$, $\bar{e_t}$ the average error of compensation of all CCs of image $I_t$, and $\sigma_t$ the standard deviation of the error. A second motion estimation is carried out with only nonmoving star CCs in order to obtain a more accurate motion estimation and a more robust classification. Finally, piece-wise tracking is carried out by extending the ($t+0,t+1$) matching with ($t+1,t+2$) matching to reduce the number of false-positive detections. For the present video data, the geometric mean error $e_t$ for the whole sequence is 0.91 pixels for the first estimation and 0.18 for the second one. The apparent speed varies from 3 up to 10 pixels/frame.

For the considered video sequence, FMDT was able to detect and track 100\% of the meteors  in the video that are visible to the naked eye, with only four false positives. The proposed solution was compared with a {manual} detection (where an expert watched and labeled the entire video). This {manual} detection constitutes the "ground truth" and was first able to detect 28 meteors, with meteors 4, 8, 16, 18, 20, and 29 being missed. The ground truth was then enhanced thanks to the automatic detection chain. This demonstrates the need for an automated system for meteor detection. Figure~\ref{plot:meteors_duration} shows the overlap between meteors detected automatically and those of the ground truth. We define the tracking rate $\mathcal{T}_r$ as the ratio of the  cumulative duration of the automatically detected meteors and of the  cumulative duration of the ground-truth meteors: 
\begin{equation*}
\mathcal{T}_r = \left( \sum_{m=1}^{34} \mathcal{D}_m^\text{auto-detect} \right) / \left( \sum_{m=1}^{34} \mathcal{D}_m^\text{ground truth} \right),
\end{equation*}
where $\mathcal{D}_m$ is the duration of the considered meteor $m$. In the observed video sequence, $\mathcal{T}_r = 80.4 \%$.
A video of the sequence with meteor tracking is available online\footnote{Meteor cluster sequence with highlighted detection: \\ \url{https://lip6.fr/adrien.cassagne/data/tauh/tracks.mp4}}.
For most of the meteors, the automatic detection is very close to reality. Moreover, the minimum time required for a {manual} detection is close to the full time of the video sequence while the proposed application is real-time and compatible with the cubesat power consumption constraint. Moreover, FMDT is able to leverage multi-core processor architectures through a task graph description and the use of the AFF3CT multi-threaded runtime library~\citep{Cassagne2019a,Cassagne2021, Cassagne2022b}. AFF3CT was designed for digital communication systems but is well adapted to real-time image processing.


\section{Discussion}\label{sec:discuss}

\subsection{Meteoroid properties}\label{sec:discuss:prop}
The measured cluster differential size distribution index $s=3.1$ is slightly lower than expected from a collision cascade \citep[3.5; see e.g.,][]{OBrienGreenberg2005}.
We examine how the derived $s$ value compares to other measurements.
Reanalyzing the 2016 SPE cluster, of which the parent body is unknown, \cite{Capek2022} find a shallow $s=1.85$.
Similarly, for decameter-size fragments ejected by 73P, \cite{Reach2009} found two relatively low size distributions of $s=1.84$ and $2.56$  for the smallest and the largest fragments, respectively.
However, given the suspected rocket effect involved in such an event, the physical process is presumably different from what is at play in a meteor cluster.
Measurements for comet 67P/Churyumov-Gerasimenko provide a wide range of values, depending on the comet heliocentric distance and meteoroid size range \citep[see][and references for a review]{Guttler2019}.
For sizes comparable to meteoroids responsible for visual meteors, extreme low values of $s=1.8$ were derived when the comet was at high heliocentric distance \citep{Merouane2016}.
Higher values of $s>3.5$ were found after perihelion \citep{Fulle2004,Moreno2017,Fulle2010}.
Last but not least, an extreme high value of $s=6.4$ was found by \cite{Vida2021} for a meteoroid fragmenting in the atmosphere.
Interestingly, extreme values of $s$ are derived for two drastically different meteoroid environments.
The lowest $s$ values are found when the meteoroid breaks up in interplanetary space, and very high values are found when this takes place in the Earth's atmosphere.
Whether this difference reflects the way meteoroids interact with gaseous environment is unclear, and investigating this matter would require additional work.

The value reported here is high compared to \cite{Koten2017}, but is still at the low end of all reported values.
\cite{Guttler2019} provide a review of the literature for 67P, and recall that meteoroids do not fragment in the coma.
It is worth pointing out that the size distribution changes as a function of heliocentric distance.
If this is true for all comets, and as the meteoroids of a given trail are ejected at different heliocentric distances, a meteoroid trail is therefore composed of a wide variety of meteoroid subtrail each described with a unique size distribution.
In addition, within a trail, the meteoroids are mixed because of the relatively wide range of sizes and ejection velocity vectors.
The portion of meteoroid subtrail sampled by the Earth during a meteor shower is therefore a mixture of all these size distributions.
The size distribution $s$ has a tremendous influence on the level of a shower \citep[number of meteor per unit of time, ][]{Vaubaillon2005a,Vaubaillon2005b}.
Future work is needed to quantify the influence of modeling a variable size distribution for the prediction of the meteor showers, and how this might reconcile past post-predictions with observations (e.g., the 2006 Leonids).

\subsection{Meteoroid cluster frequency}\label{sec:discuss:freq}
Meteoroids are known to fragment in the atmosphere with very high probability \citep[$90\%$; see][]{Subasinghe2016}. 
The reason for a cluster observation is (see above) an interplanetary fragmentation event.
However, this is very rarely observed: only seven clusters have been reported in the past $\sim 40$ years of meteor observations.
With the new data from the $\tau$-Herculids campaign, we find that the probability of such a cluster observation is $P \sim 5.5 \times 10^{-22}$ (see Sect. \ref{sec:cara}).
\cite{Sampson2007} points out that the assumption of a Poisson distribution might not be appropriate, and generally underestimates $P$.
However, if in the case presented here an extreme error of a factor $10^9$ is assumed, this still leads to $P\sim 10^{-13}$, showing the extreme rarity of the phenomenon.

Computing a meteor cluster observation frequency would require the consideration of the limiting magnitude as a function of time, the software detection efficiency, and the camera running efficiency, among others, for each meteor-detection camera.
As such a thorough study is out of the scope of this paper, we attempt to provide an order-of-magnitude estimate.
In the past 20 years, $\sim 2 \times 10^6$ meteors were observed by the IMO during a total effective observation time of $\sim 8 \times 10^6$ hours \citep{Molau2021}.
The EDMOND database currently\footnote{https://www.meteornews.net/edmond/edmond/edmond-database/, accessed on 28 July 2022} counts $\sim 4.6 \times 10^6$ meteors gathered between 2000 and 2016 \citep[][]{Kornos2014}.
The SonotaCo and GMN networks have recorded totals of  respectively $\sim 3.5 \times 10^5$ and $\sim 2.2 \times 10^5$ meteors over the past 14 years \citep{SonotaCo2021,Vida2021}.
During this time, only a handful of clusters were reported \citep{Koten2017}.
From our experience, meteor-detection software from video data (RMS, UFOCapture and FreeTure) is able to detect more than one meteor in a given frame.
This is enough to conclude that the occurrence of a meteor cluster happens with a frequency of less than one in a million meteors.

All reported clusters happened during a meteor shower \citep[][and this work]{Watanabe2003,Koten2017}.
Their orbits cover all possible cometary orbits: $\tau$-Herculids for Jupiter family comet(JFC)-type, Leonids for Halley type (HT), and September Perseids for long-period (LP) type.
The time between the disruption in the interplanetary space and the entry in the atmosphere was estimated to be only a few days.
This represents less than $0.3$\% of the orbital period of a JFC meteoroid.
A short heliocentric distance presumably increases the chances of self-fragmentation of meteoroids, given the higher micro-meteoroid space density, higher temperature and thermal stresses, and generally higher influence of radiation on their rotation state.

\subsection{Origin of meteoroid self-fragmentation}\label{sec:discuss:orig}
The cluster presented here became visible nearly 2 h after the expected maximum $\tau$-Herculids shower outburst caused by the trail ejected from comet 73P in 1995.
No encounter with any other trail was expected at this time.
The age of the parent meteoroids cannot be pinpointed, but given the lifetime expectancy of Jupiter family streams, this is likely to be a few hundred years at most \citep{Vaubaillon2019}.
Out of the currently eight meteor cluster detections (including this study), only \cite{PiersHawkes1993} was not related to a known meteor shower.
The extreme fragility of some cometary meteoroids \citep{Hornung2016} might explain this feature.
The often-quoted physical processes responsible for meteoroid fragmentation in interplanetary space are thermal stresses, collision, rotational outburst, and outgassing of volatile material.
The processes involved in the natural release of meteoroids from an active asteroid were described by \cite{Jewitt2015}.
\cite{Capek2022} found that thermal stress was most probably responsible for the 2016 SPE meteor cluster.

\subsection{Future application of the developed algorithm}\label{sec:discuss:algo}
In addition to a cluster detection, we present a first application of the new processing chain for meteor detection named FMDT.
This toolbox is derived from the CubeSat project Meteorix dedicated to the detection of meteors and space debris from space \citep{Rambaux19, Rambaux21_JIMO}.
This detection chain allow the real-time identification of meteors and enable autonomous selection of scientific data to be sent back to Earth from on board a CubeSat.
The full chain also contains an optical flow algorithm for accurate motion estimation.
The agreement in detections between the ``RMS'' software \citep{Vida.et.al2016,Vida2021} and the new approach proposed by our team \citep{Millet2022_Meteorix_COSPAR,Millet2022_Meteorix_WGN,Petreto2018_SIMD_GPU_OF_DASIP} allows us to test and validate the approach implemented and to increase the Technology Readiness Level to 5.
Such a tool might be used for future detection of meteors from orbiting spacecraft 
\citep[using e.g., the SPOSH camera;][]{Bouquet2014,Oberst2011}, or more generally from mobile observation platforms \citep{Vaubaillon2021}.

\section{Conclusion}

We describe and fully characterize the eighth meteor cluster ever reported.
Based on our analysis of the observation data, we conclude that the probability of a cluster meteor observation is less than one in a million observed meteors.
The derived differential size distribution index $s=3.1$ is relatively shallow.
This index varies with heliocentric distance for regular comet outgassing.
Future meteor-shower-prediction models might take this phenomenon into account for better accuracy.

We developed an open-source computer-vision-based toolbox, namely the Fast Meteor Detection Toolbox (FMDT) to detect and track meteors. 
In spite of the acquisition camera instability caused by the aircraft, it was able to detect 100 \% of the meteors that are detectable in the video with the naked eye, even those of high magnitude.

\begin{acknowledgements}
MoMet is supported by "Programme National de Planetologie" and IMCCE / Observatoire de Paris / PSL.
Airborne campaign organized and funded through RTI and supported by University of Southern Queensland.
Maxime Millet PhD grant is funded by Region Ile-de-France.
Dr. Fabian Zander is funded by the Australian Research Council through the DECRA number DE200101674. Dr. Juraj Toth was supported by ESA contract No. 4000128930  /19/NL/SC,
the Slovak Research and Development Agency grant APVV-16-0148, the Slovak Grant Agency for Science grant VEGA 1/0218/22.
\end{acknowledgements}

\bibliographystyle{aa}
\bibliography{cluster,ref_lacas}

\end{document}